\documentclass[a4paper,fleqn,usenatbib]{mnras}

\usepackage{graphicx}	
\usepackage{amsmath}	
\usepackage{amssymb}	
\usepackage[T1]{fontenc}
\usepackage{aecompl}

\title[The Massive Heartbeat $\iota$ Orionis]{The Most Massive Heartbeat: An In-depth Analysis of $\iota$ Orionis}

\author[H. Pablo et al.]{
Herbert Pablo,$^{1}$\thanks{E-mail: hpablo@astro.umontreal.ca}
N. D. Richardson,$^{2}$
J. Fuller,$^{3,4}$
J. Rowe,$^{5}$
A.F.J. Moffat,$^{1}$
\newauthor
R. Kuschnig,$^{6}$ %
A. Popowicz,$^{7}$ %
G. Handler,$^{8}$ 
C. Neiner,$^{9}$ %
A. Pigulski,$^{10}$ %
\newauthor
G. A. Wade,$^{11}$ %
W. Weiss,$^{12}$ %
B. Buysschaert,$^{9,13}$ %
T. Ramiaramanantsoa,$^{1}$ %
\newauthor
A. D. Bratcher,$^{2}$ %
C. J. Gerhartz,$^{2}$
J. J. Greco,$^{2}$ %
\newauthor
K. Hardegree-Ullman,$^{2}$ %
L. Lembryk,$^{2}$ %
W. L. Oswald$^{2}$ %
\\
$^{1}$D\'epartement de physique and Centre de Recherche en Astrophysique du Qu\'ebec (CRAQ), Universit\'e de Montr\'eal, C.P. 6128,\\ Succ.~Centre-Ville, Montr\'eal, Qu\'ebec, H3C 3J7, Canada\\
$^{2}$Ritter Observatory, Department of Physics and Astronomy, The University of Toledo, Toledo, OH 43606-3390, USA\\
$^{3}$TAPIR, Walter Burke Institute for Theoretical Physics, Mailcode 350-17, Caltech, Pasadena, CA 91125, USA\\
$^{4}$Kavli Institute for Theoretical Physics, Kohn Hall, University of California, Santa Barbara, CA 93106, USA\\
$^{5}$Institut de recherche sur les exoplan\`etes, iREx, D\'epartement de physique, Universit\'e de Montr\'eal, Montr\'eal, QC, H3C 3J7, Canada\\
$^{6}$Graz University of Technology, Institute of Communication Networks and Satellite Communications, Inffeldgasse 12, 8010 Graz, Austria\\
$^{7}$Silesian University of Technology, Institute of Automatic Control, Gliwice, Akademicka 16, Poland\\
$^{8}$Centrum Astronomiczne im.~M.\,Kopernika, Polska Akademia Nauk, Bartycka 18, 00-716 Warszawa, Poland\\
$^{9}$LESIA, Observatoire de Paris, PSL Research University, CNRS, Sorbonne Universit\'{e}s, UPMC Univ. Paris 06, Univ. Paris Diderot, Sorbonne Paris Cit\'{e}, 5 place Jules Janssen, 92195 Meudon, France\\
$^{10}$Instytut Astronomiczny, Uniwersytet Wroc{\l}awski, Kopernika 11, 51-622 Wroc{\l}aw, Poland\\
$^{11}$Department of Physics, Royal Military College of Canada, PO Box 17000, Station Forces, Kingston, Ontario, K7K\,7B4, Canada\\
$^{12}$Institut f\"ur Astrophysik, Universit\"at Wien, T\"urkenschanzstrasse 17, 1180 Wien, Austria\\
$^{13}$Instituut voor Sterrenkunde, KU Leuven, Celestijnenlaan 200D, 3001 
Leuven, Belgium \\
}
\date{Accepted XXX. Received YYY; in original form ZZZ}

\pubyear{2016}

\begin{document}
\label{firstpage}
\pagerange{\pageref{firstpage}--\pageref{lastpage}}
\maketitle

\begin{abstract}
$\iota$ Ori is a well studied massive binary consisting of an O9 III + B1 III/IV star. Due to its high eccentricity ($e=0.764$) and short orbital period ($P_{\rm orb}$ = 29.13376 d), it has been considered to be a good candidate to show evidence of tidal effects; however, none have previously been identified. Using photometry from the BRITE-Constellation space photometry mission we have confirmed the existence of tidal distortions through the presence of a heartbeat signal at periastron. We combine spectroscopic and light curve analyses to measure the masses and radii of the components, revealing $\iota$ Ori to be the most massive heartbeat system known to date. In addition, using a thorough frequency analysis, we also report the unprecedented discovery of multiple tidally induced oscillations in an O star.  The amplitudes of the pulsations allow us to empirically estimate the tidal circularization rate, yielding an effective tidal quality factor $Q \sim 4 \times 10^4$.
\end{abstract}





\section{Introduction}

The advent of long time-baseline photometry has changed the landscape of stellar astronomy. This is particularly true for the discovery of a unique class of binary stars known as heartbeat stars. This class, first discovered with the \textit{Kepler} space telescope \citep{thompson12}, can only be described as unusual, displaying variations that, if they were not strictly periodic, would likely not have been associated with binarity. These systems have two defining characteristics: sinusoidal pulsations on top of an otherwise stable light curve and ellipsoidal variations (bearing qualitative similarities to the ``normal sinus rhythm'' signal of an electrocardiogram) confined to orbital phases near periastron. 

Despite how peculiar these objects seem at first glance their behavior is mostly well understood. \cite{thompson12} identified the heartbeat variation as being caused by tidal distortions in a highly eccentric system. This variation can be modeled largely by the eccentricity, the inclination and the argument of periastron of the system. Beyond being distinct from other mechanisms of photometric variability, the heartbeat's strong dependence on inclination makes it a powerful tool for obtaining masses and radii for the individual components in a double-lined spectroscopic binary, even in the absence of eclipses.

What makes heartbeat systems even more valuable, is the presence of clear tidally excited oscillations (TEOs). Interactions between binary components is essential to our understanding of how such systems evolve. However, though these oscillations were postulated as early as \cite{cowl41} they were not observed until much later in the system HD 174884 \citep{maceroni09}. They were definitively confirmed with the discovery of the heartbeat star KOI 54 \citep{welsh11, fuller12, burkart:12}. Since then, these oscillations have been found in several eccentric systems \citep[see][and references therein]{shporer:16}. Additionally, we have learned that these tides not only induce pulsations, but also affect existing ones \citep{hamb13}. TEOs have even provided a way to identify the geometry of modes \citep{guo16, oleary14}.

The heartbeat phenomenon has only been observed in lower-mass stars (mainly A and F type stars). It should, however, extend to higher masses as virtually all massive stars begin their lives as binaries \citep{sana12, sana14, aldoretta15}. Additionally, due to their short life spans relative to low-mass stars, circularisation of massive binaries is often observed to be still in progress, leading to many systems which should still have high eccentricities. The lack of heartbeat systems, therefore, is likely due to observational bias as only the \textit{Kepler} and \textit{CoRoT} missions have had the continuity and sensitivity necessary to uncover these objects and their catalogs contain few B stars and only 6 O stars (all observed by \textit{CoRoT}).


This is unfortunate as massive star evolution, and O star evolution in particular, would benefit a great deal from the study of hearbeat systems. Binary interactions in massive stars are so common that mergers happen around 24 \% of the time \citep{sana12}. This means that our only hope of studying such systems where interactions are not taking place is when the period is long, and consequently eclipses, which allow for determination of masses and radii, are rare. Heartbeats therefore are a useful avenue of approach. Moreover, they come with the added benefit of pulsations- something so rarely seen in O stars that only six photometric pulsators have been confirmed \citep[see][and references therein]{bram15, pablo15}. 

The nanosatellite mission BRITE-Constellation is well-suited for just such a project, allowing for long time-baseline, high-precision photometry of the brightest stars in the sky, which also tend to be massive or luminous stars \citep{BriteI}. In its commissioning field it was found that the highly eccentric massive binary, $\iota$ Ori is in fact a heartbeat star. 
In this paper we will give a brief history of this unique system in Sect.~\ref{sect:iota} followed by a summary of the observations  in Sect.~\ref{sect:obs}. Then we will examine the determination of fundamental parameters through an orbital solution (Sect.~\ref{sect:binary}), the frequency analysis and modeling of TEOs in Sect.~\ref{sect:freq_analysis}, discuss the results obtained for this system (Sect.~\ref{sect:discuss}), and finally present conclusions  (Sect.~\ref{sect:conclusions}).   

\section{$\iota$ Orionis}   
\label{sect:iota}

$\iota$ Ori was first discovered to be a spectroscopic binary by \cite{frost03} with  the first orbital solution,  including its extreme eccentricity ($e=0.764$) coming from \cite{plasket08}. It consists of an O9 III star whose spectral type has been well established and a B1 III-IV  companion \citep{bag01} with an orbital period $P_{\rm orb}$ = 29.13376(17) d ($f_{\rm orb}$ = 0.0343244(2) d$^{-1}$) \citep{march00}. After its discovery, the system was largely ignored, excepting for occasional refinements of the orbital solution until \cite{stick87} found what was thought to be a grazing eclipse which led to the determination of masses and radii for the components. Given these parameters the components should have been close enough at periastron for tidal effects to be apparent using available instrumentation. 

This led to papers looking for colliding winds \citep{winds} and tidally induced pulsations \citep{pulsations}. However,  no evidence of tidal effects was ever found. The reason became apparent when \cite{march00}, showed that the photometric variability seen in $\iota$ Ori was not due to an eclipse. However, the amplitude of this signal relative to the noise was too low to confirm the source of variability, although tidal distortion was suspected. Due to the strong variations in the spectrum as a function of orbital phase, the temperature \citep[ $T_{\rm eff,p}\approx$~32500 K, $T_{\rm eff,s}\approx$~28000 K:][]{march00} and projected equatorial velocity \citep[$v_{\rm p} \sin i \approx$ 120 km s$^{-1}$, $v_{\rm s} \sin i \approx$ 75 km s$^{-1}$][]{gies96, march00} have been difficult to accurately determine. As a final note, evolutionary models of the two components led to speculation that this system had not co-evolved and was instead a product of a binary-binary collision \citep{bag01}. 
\section{Observations}
\label{sect:obs}
\subsection{BRITE Photometry}

All photometric observations were taking using the BRITE-Constellation, a network of nanosatellites designed to continuously monitor the brightest stars in the sky \citep{BriteI}. As the first astronomical mission using nanosatellites, data reduction is a significant effort. All light curves are processed from raw images to light curves through the procedure outlined in Popowicz et al. (2016, in prep). However, this pipeline provides only raw flux measurements and leaves all post light curve processing to the user. 

Decorrelating the extracted data requires some knowledge of the types of issues facing the mission and the instrumental signals present. This is discussed in detail by \cite{BriteII} but some of the main factors  are the lack of on-board cooling as well as minimal radiation shielding. The photometric reductions used in this paper loosely follow the typical outlier rejection and decorrelation methods described in \cite{pigu16}. Additionally, we also correct for flux changes due to change in point spread function (PSF) shape as a function of temperature, as outlined by \cite{bram16} and Buysschaert et al. (in prep). 

\begin{table}
\caption{$\iota$ Ori Photometric Observations. The first two capital letters of the satellite moniker represent the name, UniBRITE (UB), BRITE Austria (BA), BRITE Heweliusz (BH), BRITE Lem (BL) and BRITE Toronto (BT),  while the last lower case letter represents the filter, red (r) and blue (b). The quoted error is per satellite orbital mean parts per thousand (ppt).}\label{tab:briteobs}
\begin{center}
	\begin{tabular}{ | l | c | c | c | } 
	Field Name & Satellite & Duration (d) & RMS error (ppt) \\
	\hline
	\hline
	Orion I &  &  &  \\
	\hline
            & UBr & 130 & 1.30 \\
            & BAb & 105 & 1.01 \\
    Orion II &  &  &  \\        
    \hline
			& BHr & 110 & 0.62 \\
			& BTr & 58  & 0.81 \\
			& BAb & 45  & 1.30 \\
			& BLb & 91  & 1.17 \\
	\hline		            
	\end{tabular}
\end{center}
\end{table}

\begin{figure*}
\includegraphics[scale=0.45]{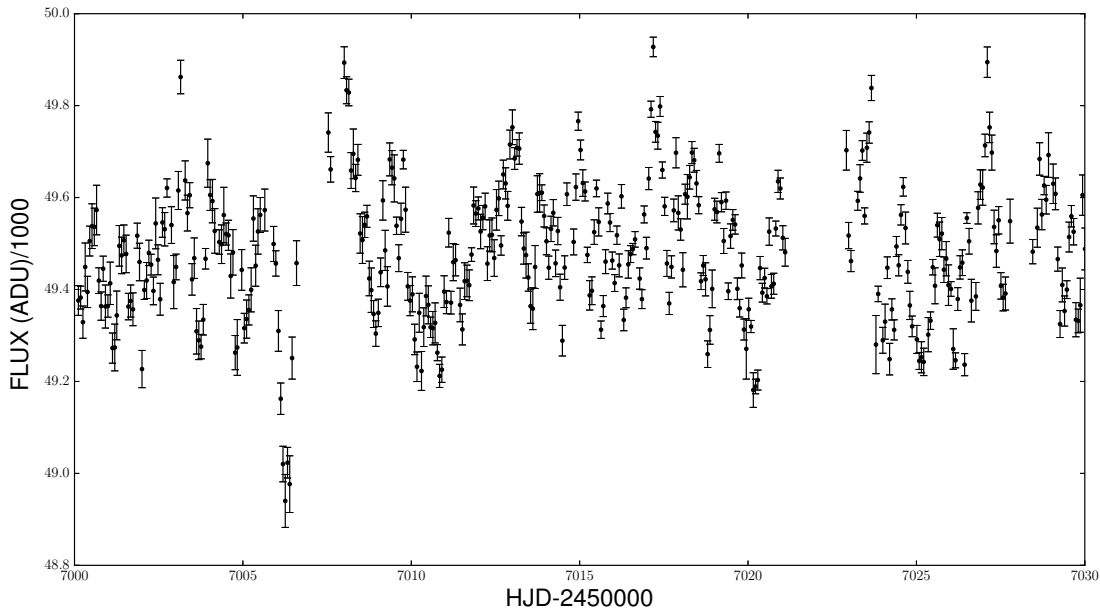}
\caption{Subset of photometry taken by BHr during the Orion II observation. This data has been cleaned of instrumental effects. A heartbeat event is visible at 2457009 d. }\label{fig:lc-time}
\end{figure*}

The observations of $\iota$ Ori detailed in Table \ref{tab:briteobs} were taken during two pointings between 2013 and 2015 with data being taken over a span of 9 months during that time. The Orion I pointing is significantly shorter and often of poorer photometric precision as it was the commissioning field for the BRITE-constellation project and much has improved in Orion II including the addition of 3 new satellites (see Figure~\ref{fig:lc-time} for an example of the photometric precision). The photometry used for analysis has been binned on the satellites orbital period for greater precision and the quoted error is the RMS error per orbit. Additionally, in cases where observations were taken from multiple satellites of the same filter within a given dataset these data were combined to form one red and one blue light curve.

For the binary fit (see Sect.~\ref{sect:binary}) virtually all the data were utilized as the amplitude of the heartbeat effect is quite large with respect to the size of the errors. However, for identifying and fitting frequencies we were significantly more selective. This included removing small chunks, on the order of a couple of days, from the beginning of certain setup files\footnote{data coming from the BRITE satellites are divided into setup files, each of which have slightly different observational parameters. This is often due to an inability to achieve fine-pointing. See \cite{BriteII} for more information.} where poor pointing accuracy led to poorer photometric quality. The most significant losses were the BAb observations in Orion II, whose errors were too large to offset the gain in frequency resolution allowed by keeping them. 

\subsection{Spectroscopy}

We collected 11 high-resolution spectra of $\iota$ Ori between 2015 November 3 and 2016
April 25 with the 1.06 m Ritter Observatory telescope and a fiber-fed echelle
spectrograph. The echelle spectrograph records spectra with a resolving power of $R$ = 26,000.  The detector is a Spectral Instruments 600 Series camera, with a 4096 $\times$ 4096 CCD with 15 $\mu$m pixels, allowing spectral coverage within 4300--7000 \AA~ across 21 orders, with some order overlap in the blue. Unfortunately, the blue range of the spectra is quite noisy due to the combination of fiber losses and instrumental response.

In general, one spectrum was obtained on each night of observation, but occasionally several spectra
were obtained in a single night. We treated the consecutive observations as
independent measurements, and the agreement in measurements was within the formal
errors. For $\iota$ Ori, the best line in these spectra to measure was the He I $\lambda$5876 line, which typically had a S/N $\gtrsim 100$. This line allowed
us to measure the two component velocities with a double-Gaussian fit, with velocity
errors for each of the two stars on the order of $\sim 5-10$ km s$^{-1}$. Other lines
towards the blue suffered from low S/N, whereas H$\alpha$ had higher S/N, but was
hard to measure due to the large intrinsic line width of H$\alpha$.

In addition to these measurements, we also used the 24 velocities reported by \cite{march00}. These augment our time baseline substantially in addition to increasing our orbital phase coverage.

\section{Binary Solution}
\label{sect:binary}

\subsection{Light Curve Cleaning}

Typically, stellar pulsations add non-coherent signal to the binary variation when phase folded on the orbital period.  However, this is no longer valid for $\iota$ Ori as the TEOs (which we discuss in detail in Sect.~\ref{sect:freq_analysis}) depend on the orbital period.  As such, their coherent signal with the orbit, which can be up to 10\% of the heartbeat effect itself, hinders our ability to derive an accurate orbital solution.  This necessitates the removal of these TEOs before we can begin our analysis of the binary system itself.  

While it would be possible to use the pre-whitened light curves obtained from our frequency analysis (see Sect.~\ref{sect:freq_analysis}), we chose to not do this because of the uncertainty associated with the frequencies, especially in the Orion I dataset. Instead, we fixed the frequency value to that of the corresponding orbital harmonic, allowing only phase and amplitude to vary. In choosing which frequencies to fit, we looked for both strength and stability across both datasets. For this reason we chose 4 frequencies believed to be TEOs ($f_2$, $f_3$, $f_4$, $f_6$ given in Table~\ref{tab:freq}). The only notable TEO not included is $f_1$. This is largely due to the fact that it does not appear strongly in the phased data, and its status as an orbital harmonic is not clear (see Sect.~\ref{sect:discuss}). Additionally, its period is long in comparison to the length of the heartbeat and as such it will not have a significant impact on the shape of the heartbeat.

For each observation and filter the data were phase folded on the orbital period. The region around the heartbeat ($\approx$ 0.2 in phase) was then removed so that it would have no effect on the amplitudes and phases of the pulsations when fit. The phased curve was then fit with a four sinusoid model using the aforementioned restrictions and a least squares minimization. The values obtained were used to create a model light curve which was then subtracted giving us a cleaned light curve. In the case of $\iota$ Ori the changes to the heartbeat were minimal, but the procedure still serves to reduce the scatter in the light curve.   

\subsection{Simulation and Fitting}

Our binary analysis consists of modeling radial velocity curves for each of the two components as well as two light curves, one in each of the BRITE filters (each a combination of OrionI and OrionII data). We are able to create simultaneous simulations for all four datasets using PHOEBE \citep{phoebe}. First we created a preliminary fit by trial and error starting with orbital parameters from \cite{march00} and getting a rough idea of the stellar parameters. These values were used to initialize our Monte Carlo Markov Chains (MCMC). MCMC probes the probability space outlined by the parameters being fit, identifying degeneracies as well as determining the extent of the global minimum for error calculation. Our MCMC implementation is achieved through the Python package {\tt emcee} \citep{mcmc}.   

Though there are well over 30 different parameters which are necessary for a complete orbital solution, MCMC is costly in time so we were careful in the choice of fit parameters. The orbital period ($P_{\rm{orb}}$) was sufficiently precise from \cite{march00} that BRITE data taken up to 15 years later phased with no noticeable phase offset. The known apsidal motion ({$\dot\omega$}) of the system is not of interest to us and also requires the fitting of $T_{0}$, the phase zero point (defined at periastron passage), so we chose not to fit these parameters and instead allow for an orbital phase shift. Furthermore as the two BRITE bandpasses do not provide enough color information to fit both temperatures we fix the value of the primary. Since the literature does not agree on a temperature and the lines change as a function of phase we take the canonical value for an O9 III star from \cite{martins05} of 31000 K. We also do not fit the masses as they can be derived, through Kepler's third law, from parameters we did fit. These parameters, 14 in total, can all be found in Table~\ref{tab:params}, the only exceptions being the phase shift and passband luminosities, which were included largely for normalization purposes and provide no quantifiable information about the system. Finally, though the limb darkening coefficients (from the logarithmic limb darkening law) were not fit, they were interpolated from precomputed tables at each iteration to correspond to the parameters of the current model.

\begin{table*}
\begin{center}
\caption{$\iota$ Ori Fundamental Parameters compared with those of Marchenk et al. (2000)}\label{tab:params}
	\begin{tabular}{r c c c c c}
	&	&  &  & \multicolumn{2}{c}{Marchenko et al. (2000)}  \\
	& Parameter & Primary   &   Secondary & Primary & Secondary \\
	\hline
	\vspace{2mm}
	& $P_{\rm orb} (d)$ & \multicolumn{2}{c}{29.13376 (fixed) } & \multicolumn{2}{c}{29.13376 (fixed)} \\
	& $T_{0} (\rm{HJD}-2450000)$ & \multicolumn{2}{c}{1121.658 (fixed)} & \multicolumn{2}{c}{1121.658 $\pm$ 0.046} \\
	& $i~(^\circ)$  & \multicolumn{2}{c}{$62.86_{-0.14}^{+0.17} $ } & \multicolumn{2}{c}{--} \\
	\vspace{2mm}
	& $\omega~(^\circ)$ & \multicolumn{2}{c}{$122.15_{-0.11}^{+0.11}$} & \multicolumn{2}{c}{$130.0 \pm 2.1$} \\
	\vspace{2mm}
	& $e$ & \multicolumn{2}{c}{$0.7452_{-0.0014}^{+0.0010} $} & \multicolumn{2}{c}{$0.764 \pm 0.007$} \\
	\vspace{2mm}
	& $q$ & \multicolumn{2}{c}{$0.5798_{-0.0084}^{+0.0077} $} & \multicolumn{2}{c}{$0.571 \pm 0.025 $} \\
	\vspace{2mm}
	& $a~(R_{\odot})$ &  \multicolumn{2}{c}{$132.32_{-0.96}^{+1.01}$} & \multicolumn{2}{c}{--}\\
	\vspace{2mm}
	& $v_{\gamma}~(km s^{-1})$ &  \multicolumn{2}{c}{$32.02_{-0.32}^{+0.30}$} & $31.3 \pm 1.2$ & $20.4 \pm 2.1$ \\
	\vspace{2mm}
	& $T_{\rm{eff}}~(K)$ & $31000$ (fixed) & $18319_{-758}^{+531}$ & -- & -- \\
	\vspace{2mm}
	& $R~(R_{\odot})$ & $9.10_{-0.10}^{+0.12} R_{\odot}$ & $4.94_{-0.23}^{+0.16}$ & -- & -- \\	
	\vspace{2mm}
	& $f$ & $14.86_{-0.23}^{0.34} $ & $28.14_{-2.017}^{+2.78} $ & -- & --\\
	\vspace{2mm}
	& $M~(M_{\odot})$ & $23.18_{-0.53}^{+0.57} M_{\odot}$ & $13.44_{-0.30}^{+0.30} $ & -- & \vspace{1mm}--\\
	\hline
	\vspace{2mm}
\end{tabular}
\end{center}
\end{table*}

\begin{figure*}
\begin{center}
\includegraphics[scale=0.25]{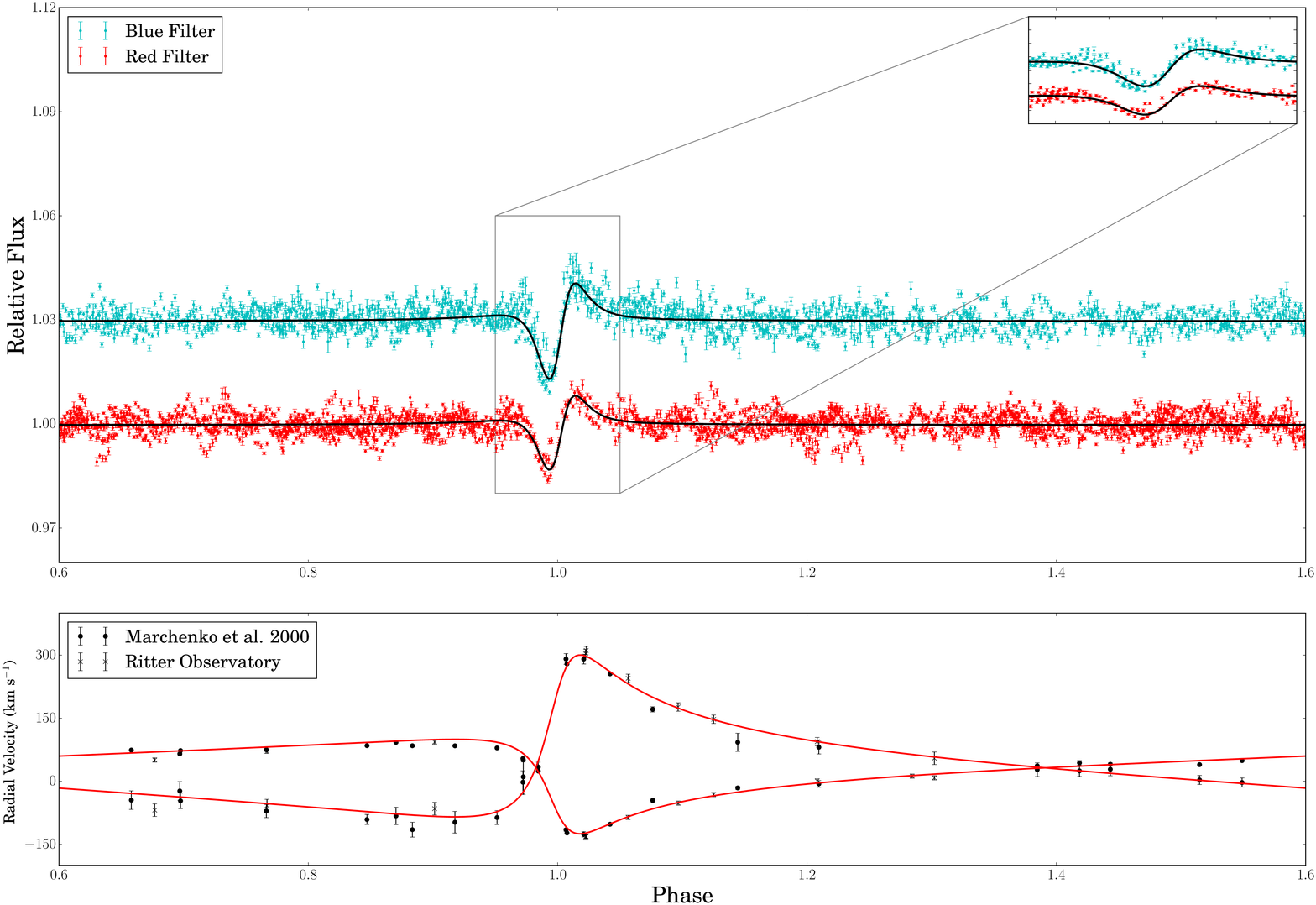}
\caption{Binary solution of $\iota$ Ori. In the top panel are the phase-folded red filter light curve (red) and blue filter light curve (cyan) overlaid with the PHOEBE fit (black). The data shown here has not been cleaned in order to show the TEOS present in both light curves in the top panel, most notably around phase=0.9. The blue filter light curve has been artificially shifted in flux to facilitatethe  display both light curves. The bottom panel shows the radial velocity data from Marchenko et al. (2000) (black x) and Ritter Observatory (black dots) overlaid with the PHOEBE fit in red.}\label{fig:binfit}
\end{center}
\end{figure*}

With a large number of parameters, a correspondingly large number of independent chains, known as walkers, should be used to efficiently explore the entire parameter space. In our case 50 walkers were chosen. As certain parameters were initially not well constrained it took over 7500 iterations, each with 50 walkers, to reach a point where the values of all parameters were stable. From there, the system was allowed to iterate 8000 more times. At this point all the parameters passed the Gelman-Rubin criterion for convergence \citep{gelman92}, when the in-chain variance is within 10\% of the variance between chains, and there were enough total iterations for valid statistics. Simulated data derived from these parameters are shown in Fig~\ref{fig:binfit}. 

The individual values quoted in Table \ref{tab:params} are the 50th percentile values for each parameter given with 2$\sigma$ error bars. The orbital elements also agree with those that \cite{march00} found. While the differences are not always within errors it is likely that those of \cite{march00} are slightly underestimated. Furthermore, the stellar parameters are largely consistent with what one would expect given the spectral types and the information known with two key differences. First, the companion  temperature is significantly lower than that of normal B1 stars. Second are the synchronicity parameters ($f$), which imply rotational periods of $\approx$ 2 and $\approx$ 1 day for the primary and secondary component respectively. While these periods equate to speeds far from critical velocity, they are faster than expected from $v \sin i$ values and what is used in the models in Sect.~\ref{sect:freq_model}. These issues will be discussed further in Sect~\ref{sect:discuss}.

\section{Frequency Analysis}
\label{sect:freq_analysis}

\subsection{Frequency Determination}
\begin{figure*}
\includegraphics[scale=0.5]{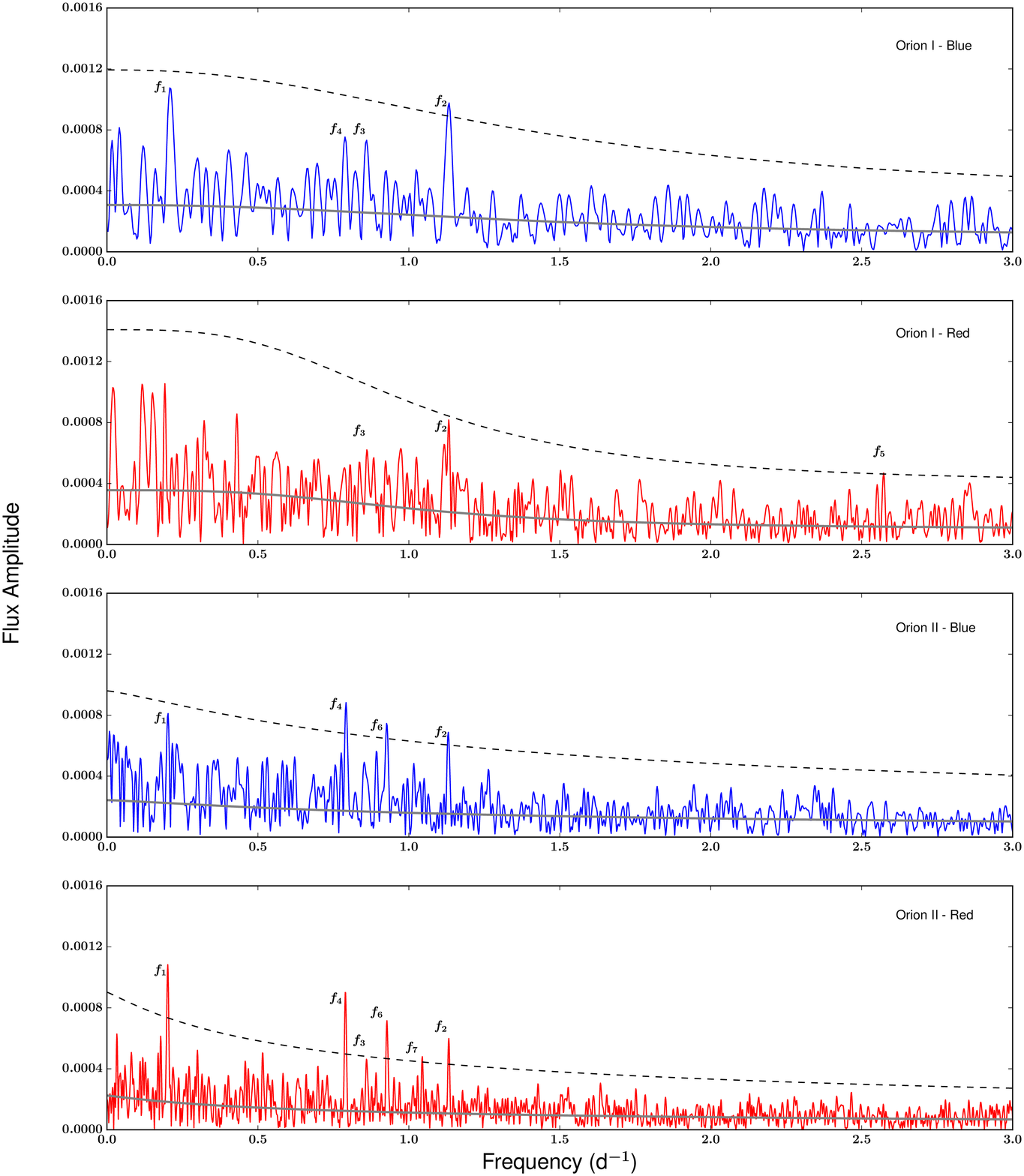}
\caption{Discrete Fourier spectrum of the BRITE data, binned on orbital phase, for Orion I blue (top) and red (second from top), and Orion II blue (third from top), and red (bottom). The dashed line in each graph denotes the 3.6 $\sigma$ signifance level, though some peak values will change during the process of pre-whitening other signals. The grey line in each graph denotes the noise floor.}\label{fig:fts}
\end{figure*}

As this data set is both large and complex, a significant amount of preparation was necessary before beginning frequency analysis. As the Orion I and Orion II campaigns were separated by a year, it was necessary to split the analysis of individual oscillation frequencies by observation as well as by color. Next, the binary modulation which is discussed in detail in Sect.~\ref{sect:binary} was removed separately from each dataset by subtracting a model of the binary variation, consistent with the parameters shown in Table~\ref{tab:params}, in the appropriate color. Finally it was important to determine an adequate significance threshold. Since the pulsations seen in $\iota$ Ori exhibit frequencies that are largely less than 1 d$^{-1}$ (see Table~\ref{tab:freq}) red noise can, and does, play a significant role (see Fig.~\ref{fig:fts}) and thus the noise floor must be fit before continuing. 

Using the procedure outlined by \cite{noise_floor} the background noise level was calculated by fitting power density ($PD$) as a function of frequency in log-log space using the following form:

\begin{equation}
PD = \frac{A}{1+(\tau f)^{\gamma}} + c\rm{,} 
\end{equation}    
where $c$ is the constant white noise, $A$ is the amplitude, $\tau$ is the characteristic timescale associated with the signal, $f$ is the frequency, and $\gamma$ is the power index. While it is not uncommon to use two or more of these semi-Lorentzians to fit the noise floor,  one was sufficient in all of our datasets. To ensure that this noise floor was not altered by power from real signals, peaks that were significantly higher than the median level were each fit simultaneously with the noise floor using Lorentzians. The noise floor as well as the significance threshold can be seen clearly in Figure~\ref{fig:fts}. While the noise floor is well defined, we were unable to gain any extra information from the derived parameters. Many had large errors, but even $\gamma$, which was well constrained for each dataset, varied significantly between both the epochs and colors of observations. One possible explanation for this disparity in the noise floor is some combination of both stellar and instrumental signal which is difficult to disentangle, especially as the Orion II datasets contain data from two different satellites.

With this done it was now possible to set our detection threshold. Following the procedure laid out by \cite{sigthresh} we define a false alarm probability, $FAP$, that in a series of $N$ frequency bins that at least one peak will be $m$ times above the noise given Gaussian statistics:

\begin{equation}
FAP_{N}(m) = 1 - (1 - e^{-m})^{pN}\rm{,}
\end{equation}
where $p$ is an empirical term necessary when oversampling the Fourier transform. For an oversampling of 5, which we use, $p$ is equal to 2.8 \citep{sigthresh}. The way our FAP is defined, the lower  its value the more likely that a given peak is real. We then pick a significance threshold which equates to a $FAP$ of 0.05 \%.  

With our limits determined we now apply a standard pre-whitening procedure, iteratively fitting sinusoids to significant peaks found in the Fourier transform. Each time a new peak is found, the light curve  is fit in combination with all other frequencies and the residuals are used to determine if any significant frequencies remain. This was carried out using the Period04 software package \citep{period04}. All signals above the detection threshold, as well as peaks near the threshold and found in multiple datasets are included in our fit and given in Table~\ref{tab:freq}. Additionally, as most significant peaks are also orbital harmonics, those peaks which were above the detection threshold in the phase folded light curve ($FAP_{ph}$) were also included.

\begin{table*}
\caption{$\iota$ Ori pulsation frequencies. Column 1 and 2 are the observation and filter in which the frequencies were found. Column 3 gives the frequency number while column 4 gives the corresponding orbital harmonic associated with this frequency. Columns 4, 5, and 6 give the frequency, amplitude and phase (phased to periastron) respectively with errors. The frequency errors are quoted as the resolution of the dataset (1/baseline) and the errors in phase and amplitude are calculated through a Monte Carlo simulation. Column 7 gives the $FAP$, in percent, of the frequency in the time domain, while column 8 gives the  $FAP$ value of the same frequency from the light curve folded on the binary's orbital period.}

\begin{center}
	\begin{tabular}{l l c c c c c c c c } 
	ID & Filter & Frequency & Orb. harmonic & $f$ (d$^{-1}$) & A (ppt) & Phase & $FAP$ & $FAP_{ph}$ \\ 
Orion I & blue & & & & & & & &\\
\hline
	& & $f_1$ & 6 & 0.211$\pm$ 0.017 & 1.06$\pm$ 0.48 & 0.324$\pm$ 0.206 & 0.625 & $1.43\times10^{-9}$ \\
	& & $f_2$ & 33 & 1.132$\pm$ 0.017 & 0.96$\pm$ 0.4 & 0.867$\pm$ 0.215 & $2.41\times10^{-3}$ & $1.91\times10^{-9}$ \\
	& & $f_3$ & 25 & 0.859$\pm$ 0.017 & 0.75$\pm$ 0.43 & 0.149$\pm$ 0.166 & 22.6 & $1.7\times10^{-4}$ \\
	& & $f_4$ & 23 & 0.789$\pm$ 0.017 & 0.7$\pm$ 0.32 & 0.315$\pm$ 0.44 & 61.7 & 0.05 \\
Orion I  & red & & & & & & & &\\
\hline    
	& & $f_2$ & 33 & 1.1324$\pm$ 0.0097 & 0.82$\pm$ 0.25 & 0.648$\pm$ 0.132 & 0.084 & $4.72\times10^{-8}$ \\
	& & $f_3$ & 25 & 0.8620$\pm$ 0.0097 & 0.62$\pm$ 0.18 & 0.455$\pm$ 0.063 & $> 99$ & $6.67\times10^{-3}$  \\
	& & $f_5$ & 75 & 2.5724$\pm$ 0.0097 & 0.45$\pm$ 0.15 & 0.205$\pm$ 0.081 & 0.053 & 0.048 \\
Orion II & blue & & & & & & & &\\
\hline
	& & $f_4$ & 23 & 0.792$\pm$ 0.010 & 0.97$\pm$ 0.13 & 0.382$\pm$ 0.02 &  $3.27\times10^{-9}$ & $1.83\times10^{-10}$ \\
	& & $f_1$ & 6 & 0.202$\pm$ 0.010 & 0.85$\pm$ 0.35 & 0.985$\pm$ 0.215 & 0.10 & 0.10 \\
	& & $f_6$ & 27 & 0.923$\pm$ 0.010 & 0.78$\pm$ 0.13 & 0.869$\pm$ 0.028 & $3.37\times10^{-5}$ & $2.98\times10^{-6}$ \\
	&  & $f_2$ & 33 & 1.13$\pm$ 0.01 & 0.7$\pm$ 0.22 & 0.09$\pm$ 0.158 & $1.26\times10^{-4}$ & $1.26\times10^{-5}$ \\

Orion II  & red & & & & & & & &\\
\hline
	& & $f_1$ & 6.0 & 0.2016$\pm$ 0.0059 & 1.07$\pm$ 0.4 & 0.553$\pm$ 0.149 & $4.08\times10^{-10}$ &   $\mathbf{< 1\times10^{-10}}$ \\
	& & $f_4$ & 23 & 0.7895$\pm$ 0.0059 & 0.92$\pm$ 0.09 & 0.734$\pm$ 0.015 & $< 1\times10^{-10}$ & $< 1\times10^{-10}$ \\
	& & $f_6$ & 27 & 0.9271$\pm$ 0.0059 & 0.66$\pm$ 0.29 & 0.504$\pm$ 0.242 & $1.93\times10^{-9}$ & $8.51\times10^{-8}$ \\
	& & $f_2$ & 33 & 1.1325$\pm$ 0.0059 & 0.58$\pm$ 0.09 & 0.211$\pm$ 0.026 & $6.70\times10^{-8}$ & $1.77\times10^{-10}$ \\
	& & $f_7$ &  & 1.0445$\pm$ 0.0059 & 0.45$\pm$ 0.09 & 0.772$\pm$ 0.035 & 0.021 & -- \\
	& & $f_3$ & 25 & 0.8597$\pm$ 0.0059 & 0.44$\pm$ 0.09 & 0.452$\pm$ 0.035 & 0.338 & $3.71\times10^{-3}$ \\

	\end{tabular}
	\label{tab:freq}
\end{center}
\end{table*}

One of the more obvious findings is that virtually all of the frequencies are orbital harmonics including $f_1$, $f_2$, $f_3$, $f_4$, and $f_6$. This is often seen in heartbeat stars and will be discussed in more detail in Sect.~\ref{sect:freq_model}. However, the relatively long orbital period of $\iota$ Ori, $P_{\rm orb}$ = 29.13376(17) d ($f_{\rm orb}$ = 0.0343244(2) d$^{-1}$), relative even to the length of the BRITE data ($\approx$ 6 months in the best case) can give us rather large frequency uncertainties. This is mostly only a problem for Orion I, but given that pulsations are rare in O stars, and TEOs are thus far unprecedented, we must be certain of our findings. 

Fortunately, TEOs are extremely stable in frequency and phase due to the steady tidal forcing produced by the orbit. They are usually also stable in amplitude, although in a few cases significant changes have been detected over time spans of years \citep{oleary:14,guo16}. This means that when the data are folded on the orbital period, the tidal oscillations not only persist, but tend to stand out more strongly as most other signals will become incoherent in the phased domain. We take advantage of this fact by taking the Fourier transform of the phased data and multiplying the frequency by the orbital period to convert frequencies from cycles per phase into d$^{-1}$. From here we can follow the same procedure described above to determine the noise level and significance threshold. The frequency resolution in this case is always $P_{\rm orb}^{-1}$, as the phased data are always exactly one orbit in length. While this is too poor to give accurate values for each frequency, it does give us confirmation that a given peak is indeed an orbital harmonic. Additionally, for modes which are only mildly significant in the time domain, this technique can often provide a second, much stronger, detection (see $\sigma_{ph}$ column in Table~\ref{tab:freq}) confirming their validity. 

As a final test, we again take advantage of the stability of tidally induced oscillations by combining both the Orion I and Orion II datasets together and achieve much higher frequency resolution. The downside of combining data with such large gaps is that aliasing makes it very difficult to determine which peaks are real. This was no less true in our case as the yearly alias was extremely strong. In the blue filter this aliasing effect was too strong for there to be any significant benefit. However, in the red filter, where the data were significantly more continuous, we were able to do this analysis effectively. Here we recovered all five suspected harmonics above the significance threshold and all but $f_{1}$ are exact multiples of the orbital frequency to a resolution of 0.0024 d$^{-1}$. 

\begin{figure}
\label{fig:spec-win}
\includegraphics[scale=0.4]{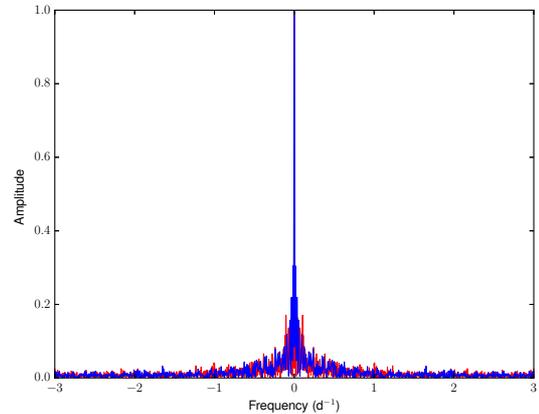}
\caption{Spectral window for the Orion II data in the blue filter (blue) and the red filter (red).}
\end{figure}

Of the suspected TEOs, there are clear instances of phase differences between observations as seen in $f_{3}$. However, this can be explained by  the fact that these frequencies - while resulting from the same orbital harmonic - have slightly different derived values which can influence the phase. Indeed when we enforce the frequency to be exactly at the closest orbital harmonic, the phase variation between observations was almost always within errors. The one tidal frequency about which there are still some inconsistencies is $f_1$. It is well below the significance threshold in Orion I, despite its prominence in the blue filter. This along with several other similarly sized signals near $f_1$ in the Orion I dataset are likely the reason it does not appear as an orbital harmonic in the combined data. Still, because it is not well constrained we must explore it further.   One possible explanation is that it is an alias peak as $f_1$+$f_4$ $\approx$ 1  d$^{-1}$. However, there is no significant power at 1 d$^{-1}$ in the Fourier transform or in the spectral window (see Figure~\ref{fig:spec-win}) so this does not seem likely. Furthermore, the phase and amplitude of $f_1$ are extremely stable in the blue filter when the frequency is fixed to the closest orbital harmonic (6$f_{orb}$). At this point we can neither prove or disprove its existence as TEO, though it will be discussed further in Sect.~\ref{sect:freq_model} and Sect.~\ref{sect:discuss}. 

The other two significant frequencies $f_5$ and $f_7$ must be evaluated separately. The source for $f_7$ is unclear as it is seen only in the red filter of Orion II. It appears almost halfway between two orbital harmonics meaning it is not due to tidal interaction. While it is close to 1 d$^{-1}$, the resolution is sufficient to rule out an alias peak. Since it is only present in one dataset though, it is unlikely to be stellar in nature and is more than likely an artifact.   

$f_5$ must be considered more carefully as it is not only within errors an orbital harmonic, but also has a strong value of $\sigma_{ph}$. Normally this would be sufficient to consider this peak as belonging to one of the binary components. However, this frequency appears only in the red filter of the Orion I dataset and its placement is unusual. All other significant peaks appear in the range of 0.79 to 1.13 d$^{-1}$ save $f_1$, but $f_{1}$'s existence is confirmed by its presence in multiple datasets. The generation of these modes will be discussed in more detail in Sect.~\ref{sect:discuss}, but it is clear from the data that there is a favored range of frequencies. It is also exactly 3$\times f_3$ making it a possible sampling alias.  These facts alone are not enough to discount $f5$, but in combination with the fact that it does not recur and is only visible in one filter, its existence is suspect. It is more likely that this is an alias peak or an possibly instrumental signal.

\subsection{Tidally Excited Oscillations}
\label{sect:freq_model}

To understand the TEOs of $\iota$~Ori, we constructed stellar models with parameters that approximately match those listed in Table \ref{tab:params}, using the MESA stellar evolution code \citep{paxton:11,paxton:13,paxton:15}. We then calculated stellar oscillation modes of our models using the non-adiabatic version of the GYRE pulsation code \citep{townsend:13}, accounting for rotation using the traditional approximation. Finally we calculated the luminosity perturbations of the modes when forced by the tidal potential of the other star. We consider only $l=2$ modes, as $l>2$ modes are much more weakly excited by the tidal potential which scales as $(R/a)^{(l+1)}$. Our method closely follows that outlined by Fuller (in prep), and that of \cite{fuller12} and \cite{burkart:12}. 

When modeling the stars, we needed to make some assumptions about the stellar spins. A rough estimate of the stellar synchronization time scale can be found from a parameterized tidal model. We use equation 11 of \cite{hut:81} (ignoring the term in brackets), with a tidal lag time $\tau = P_{\rm orb}/(2 \pi Q)$, choosing $Q=4 \times 10^4$, a reasonable guess based on the value we calculate below. This yields a pseudo-synchronization timescale of order $10^4$ years, and a corresponding circularization time scale of order $10^7$ years. Therefore the system has plausibly reached a pseudo-synchronous spin state without having circularized yet, so we assume the stellar spin axes are aligned with the orbital angular momentum axis. The pseudo-synchronous spin period (which is independent of the synchronization rate) is $P_{ps} = 3.4 \, {\rm days}$, but this value depends on the tidal prescription and should be considered a rough estimate of the actual spin periods. For aligned spin and orbit, the $l=2, \, m=\pm1$ modes are not excited, and we expect the $l=2$, $m=\pm2$ and $m=0$ modes to dominate the tidal response of the star.

\begin{figure*}
\includegraphics[scale=0.4]{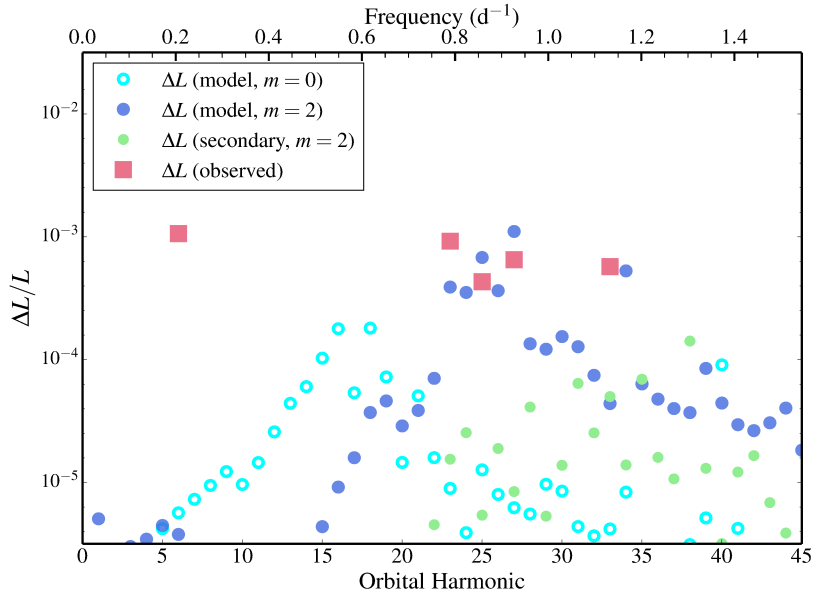}
\caption{Observed and modeled luminosity fluctuations produced by tidally excited stellar oscillations in the $\iota$~Ori system. Non-axisymmetric $l=m=2$ prograde g modes of the primary are most likely to be responsible for the observed oscillations, with the exception of the lowest frequency pulsation (see text).}\label{fig:tide}

\end{figure*}

Figure \ref{fig:tide} presents a comparison between the observed and modeled TEOs, showing the predicted luminosity fluctuations $\Delta L/L$ produced by TEOs at each orbital harmonic $N$. We plot the contribution of $|m|=2$ modes for both the primary and secondary stars, and also the contribution of axisymmetric $m=0$ modes of the primary. We find that the prograde (i.e., same direction as the orbital motion) $|m|=2$ modes of the primary are expected to dominate the TEOs, and both their predicted frequencies and amplitudes are similar to the observed modes in $\iota$~Ori. 
The smaller contribution from the secondary originates mostly from its smaller contribution to the total flux.

Figure \ref{fig:tide} adopts a spin period $P_{s} = 4.5 \, {\rm days}$ for each star, slightly longer than the pseudo-synchronous spin period, and much longer than the light curve modeling estimates from Table~\ref{tab:params}. Much faster/slower spin rates generally shift the modeled TEO frequencies
to signiﬁcantly higher/lower values than observed. Although the spin rate does not change the tidal forcing frequencies in the inertial frame, it changes which modes are resonant with this forcing. Faster spin (when
aligned with the orbit) boosts g modes to higher frequencies in the inertial frame, causing resonant modes that couple strongly with tidal forcing to be excited at larger orbital harmonics. Smaller spin or
retrograde spin has the opposite effect \citep[see][]{lai96}. Faster aligned rotation thus makes it more likely to observe higher frequency TEOs. Although we cannot use this effect to measure a precise rotation rate, we can disfavor very fast/slow rotation because these models tend to produce higher/lower frequency TEOs than observed. 

Our models cannot explain the $f_1$ pulsation near $N=6$. If this pulsation arises from an $|m|=2$ mode, it must be produced by a retrograde mode (as measured in the rotating frame of the star). Although our models include retrograde g modes, they are very weakly excited. However, our models do not include Rossby modes, which couple much more strongly to tidal forcing because of their smaller radial wavenumber (see discussion in \cite{fuller:14}). These types of modes have been recently observed in several $\gamma$-Doradus stars \citep{vanreeth:16}, and it is possible the pulsation at $N=6$ is produced by a tidally excited Rossby mode in the primary of $\iota$~Ori. Another possibility is that this pulsation arises from non-linear interactions between TEOs, which have been observed in several heartbeat stars (Fuller et al. 2012, Hambleton et al. 2013, O'Leary et al. 2014, Guo et al. 2016). The signature of non-linear interactions is combination frequencies between three modes such that $f_a \pm f_b = f_c$. Indeed, in $\iota$~Ori, $f_1 + f_6 \simeq f_2$, indicating non-linear effects could be at play. If so, we speculate that $f_1$ is a mode non-linearly excited by the interaction between TEOs at $f_2$ and $f_6$.

We can estimate the tidal dissipation rate in the $\iota$~Ori system based on the observed pulsation amplitudes. The energy dissipation rate due to each stellar oscillation mode can be calculated (see equations 11-14 of \cite{burkart14}) from a model that roughly reproduces the observed oscillations. Using the model shown in Figure \ref{fig:tide}, we calculate the rate of change of orbital energy to be $\dot{E}_{\rm orb} \approx 7 \times 10^{34} \, {\rm erg} \, {\rm s}^{-1}$. This corresponds to an orbital decay rate $t_{\rm tide} = E_{\rm orb}/\dot{E}_{\rm orb} \approx 3 \times 10^7 \, {\rm yr}$ which is compatible with the young age of the system and its high eccentricity. This can also be expressed in terms of an effective tidal quality factor $Q_{\rm tide}$, which we define (note some differences between our definition and that of \cite{hut:81}) via 
\begin{equation}
\label{Q}
\frac{\dot{E}_{\rm orb}}{E_{\rm orb}} = \frac{3 k_2}{Q_{\rm tide}} \bigg(\frac{R}{a_p}\bigg)^{\!5} \Omega \, ,
\end{equation}
where $k_2$ is the primary's Love number, $a_p = a(1-e)$ is the periastron orbital separation, and $\Omega = 2 \pi/P_{\rm orb}$ is the angular orbital frequency. For our model, we find $k_2 = 6 \times 10^{-3}$ and $Q_{\rm tide} = 4 \times 10^4$. Although our result is model dependent, to the best of our knowledge it represents the first empirical estimate of the tidal dissipation rate in an O star. We emphasize that this estimate accounts only for dissipation via tidally excited g modes, and other undetected sources of tidal dissipation could exist. Additionally, this value of $Q_{\rm tide}$ can change as a function of stellar mass, evolutionary state, orbital eccentricity, etc., and should not be interpreted as a universal constant for O type stars.



\section{Discussion}
\label{sect:discuss}

We have derived a consistent solution for $\iota$~Ori in which the tidal model matches well with the binary solution. However, our findings do raise a few questions with respect to previous works, most notably with the derived luminosity classes of the components. The mass and radius found in Sect.~\ref{sect:binary} for the secondary star are consistent with those with a spectral type between B1 and B0.5, and luminosity class  IV or V, which seem to match reasonably well with the B1 III/IV (B0.8 III/IV interpolated) spectral type from \cite{bag01}. However, the mass--radius relationship would seem to rule out luminosity class III based on current evolutionary track models \citep{nieva14}. In fact, though we cannot completely rule out class IV, we see no discernible difference from that of a class V star. 

This disparity is likely due to the difficulty in disentangling the spectra due to the large line variations of both components as a function of orbital phase \citep{march00, bag01}. The primary star also has values more consistent with a class IV star from evolutionary models given by \cite{martins05}. This is less concerning though as there are very few empirical constraints for O III stars and so the range of allowed values has not been adequately explored. We note that while our spectral types are much more acceptable for a co-evolved system than previous ones, we can neither confirm nor deny the claims non-coevolution made by \cite{bag01}. 

Another issue is the temperature of the secondary, which even taking into account errors is several thousand kelvin lower than expected from its spectral type. One reason for this may have to do with the value we used for the temperature of the primary. The primary is a known O9 star; however its mass and radius would signify something more akin to an O8 star. Even with the corresponding adjustment in temperature though, we only increase the secondary's temperature by $\approx$ 2000 K,  which is still much lower than expected. While it is possible that we have not found the global minimum in our MCMC analysis, this is also unlikely as we have rerun the analysis with various initial values of the temperature, to ensure that our result does not depend on our choice of priors. Since it is unlikely that the secondary's temperature is indeed this low, we are forced to conclude that the model is simply not very sensitive to temperature and is likely degenerate with other parameters which were not fit. 
The final issue with the binary solution is that the rotation periods derived from the synchronicity parameters are each a factor of 2 shorter than what $v \sin i$ measurements as well as our tidal oscillation models would suggest. Like with temperature we are forced to conclude that our light curve model is not very sensitive to rotation and that the effect rotation does have is likely degenerate with limb or gravity darkening. 

In addition, we do have one issue with our TEO model which is the existence of $f_1$. Though we have a possible explanation through non-linear interactions, we should also consider other possibilities for this peak. It could be a pulsation that is unrelated to the tides, though the fact that it is very close to an orbital harmonic and, in the blue filter, is extremely stable in phase and amplitude would be a rather strange coincidence. Another possibility, given its period of $\approx$  4.8 days,  is that this frequency is related to the star's rotation period. While this does make sense, the peak seems unusually stable given our limited knowledge about the timescales associated with spots in O stars. While we lean toward the idea that this is likely an orbital harmonic it is impossible to say definitively without further observations.

\section{Conclusions \& Future Work}
\label{sect:conclusions}
We have identified and obtained fundamental parameters for $\iota$~Ori, the most massive known heartbeat system. We have also discovered tidally induced pulsations for the first time ever in an O type star and confirmed that these frequencies agree well with those predicted by models. Furthermore we have shown the impact the BRITE-Constellation project has on our knowledge of massive stars. For the first time we are able to explore the asteroseismic effects of binarity in massive stars and in a system that was quite literally hidden in plain sight, including the first empirical calculation of tidal dissipation rate in an O star (see Sect.~\ref{sect:freq_model}).

Despite the unprecedented nature of our findings, this marks only the beginning of our attempts at unraveling the mysteries of this system. In addition to what has been presented here, by May of 2017 we will have two more observation sequences from BRITE from which to hopefully derive more frequencies and refine our asteroseismic models. We have also been granted time on NPOI to resolve the binary with interferometry, while efforts are also underway with the CHARA Array.  These should allow us to obtain a distance and an independent measure of the system's inclination to confirm the results reported herein. 

\section*{Acknowledgements}
Based on data collected by the BRITE Constellation satellite mission, designed, built, launched, operated and  supported by the Austrian Research Promotion Agency (FFG), the University of Vienna, the Technical University of Graz, the Canadian Space Agency (CSA), the University of  Toronto  Institute  for  Aerospace Studies (UTIAS),  the  Foundation for Polish Science \& Technology (FNiTP MNiSW), and National Science Centre (NCN). We would like to thank the computing infrastructure at Villanova for use of their cluster. NDR acknowledges postdoctoral support by the University of Toledo and by the Helen Luedtke Brooks Endowed Professorship. AFJM and HP are grateful for financial aid from NSERC (Canada) and FQRNT (Quebec).GAW acknowledges Discovery Grant support from the Natural Science and Engineering Research Council (NSERC) of Canada. AP acknowledges support from the NCN grant No. 2016/21/B/ST9/01126. JF acknowledges partial support from NSF under grant no. AST1205732 and through a Lee DuBridge Fellowship at Caltech. GH acknowledges support from the Polish NCN grant 2015/18/A/ST9/00578. The Polish participation in the BRITE project is 
secured by NCN grant 2011/01/M/ST9/05914.




\bibliographystyle{mnras}
\bibliography{biblio} 







\bsp	
\label{lastpage}
\end{document}